\documentclass[conference]{IEEEtran}
\usepackage{amsfonts}
\usepackage{times}
\usepackage{graphicx}
\usepackage{latexsym}
\usepackage{dsfont}
\usepackage{amssymb}
\usepackage{amsmath}
\usepackage{cite}
\usepackage{verbatim}

\newcommand{\figref}[1]{{Fig.}~\ref{#1}}

\def\bb0{{\mathbb{0}}}

\def\ba{{\mathbf{a}}}
\def\bb{{\mathbf{b}}}

\def\bh{{\mathbf{h}}}

\def\bm{{\mathbf{m}}}
\def\bn{{\mathbf{n}}}

\def\bp{{\mathbf{p}}}
\def\bq{{\mathbf{q}}}
\def\br{{\mathbf{r}}}
\def\bs{{\mathbf{s}}}

\def\bx{{\mathbf{x}}}
\def\by{{\mathbf{y}}}
\def\bz{{\mathbf{z}}}
\def\b0{{\mathbf{0}}}

\def\bA{{\mathbf{A}}}

\def\bF{{\mathbf{F}}}

\def\bH{{\mathbf{H}}}
\def\bI{{\mathbf{I}}}

\def\bN{{\mathbf{N}}}

\def\bP{{\mathbf{P}}}
\def\bQ{{\mathbf{Q}}}

\def\bU{{\mathbf{U}}}
\def\bV{{\mathbf{V}}}
\def\bW{{\mathbf{W}}}

\def\bY{{\mathbf{Y}}}

\def\bbC{{\mathbb{C}}}

\def\bbE{{\mathbb{E}}}

\def\bbR{{\mathbb{R}}}

\def\cD{\mathcal{D}}
\def\cE{\mathcal{E}}
\def\cF{\mathcal{F}}

\def\cH{\mathcal{H}}

\def\cL{\mathcal{L}}

\def\cW{\mathcal{W}}

\def\sf0{{\mathsf{0}}}

\usepackage{epstopdf}
\usepackage{enumerate}
\usepackage{float}
\usepackage{color}
\usepackage{makeidx}
\usepackage{bm}
\usepackage{cleveref}
\usepackage{url}
\usepackage{subfigure}
\usepackage{balance}

\newcommand{\sref}[1]{{Section}~\ref{#1}}
\DeclareMathOperator*{\argmin}{arg\,min}

\begin{document}

\title{Digital Twin Aided Compressive Sensing: \\ Enabling Site-Specific MIMO Hybrid Precoding}

\author{Hao~Luo~and~Ahmed~Alkhateeb\\School of Electrical, Computer, and Energy Engineering, Arizona State University\\Email: \{h.luo, alkhateeb\}@asu.edu}

\maketitle

\begin{abstract}	
    Compressive sensing is a promising solution for the channel estimation in multiple-input multiple-output (MIMO) systems with large antenna arrays and constrained hardware. Utilizing site-specific channel data from real-world systems, deep learning can be employed to learn the compressive sensing measurement vectors with minimum redundancy, thereby focusing sensing power on promising spatial directions of the channel. Collecting real-world channel data, however, is challenging due to the high overhead resulting from the large number of antennas and hardware constraints. In this paper, we propose leveraging a site-specific digital twin to generate synthetic channel data, which shares a similar distribution with real-world data. The synthetic data is then used to train the deep learning models for learning measurement vectors and hybrid precoder/combiner design in an end-to-end manner. We further propose a model refinement approach to fine-tune the model pre-trained on the digital twin data with a small amount of real-world data. The evaluation results show that, by training the model on the digital twin data, the learned measurement vectors can be efficiently adapted to the environment geometry, leading to high performance of hybrid precoding for real-world deployments. Moreover, the model refinement approach can enable the digital twin aided model to achieve comparable performance to the model trained on the real-world dataset with a significantly reduced amount of real-world data.

\end{abstract}

\section{Introduction}
    Multiple-input multiple-output (MIMO) systems can employ large antenna arrays to achieve high beamforming gain and enable spatial multiplexing, thereby improving the spectral efficiency. Traditionally, signal processing in the MIMO systems is performed digitally at the baseband, allowing for control over the phase and amplitude of the signals. However, as the number of antennas increases, the hardware cost and power consumption of the mixed-signal hardware restrict the number of radio frequency (RF) chains that can be utilized in practice. To overcome this challenge while enjoying the benefits brought by large antenna arrays, constraints have been imposed on the hardware architecture, leading to the development of various designs. For instance, hybrid analog/digital precoding~\cite{Ayach2014} has been proposed to reduce the number of RF chains by dividing the precoding operation into two stages: Digital baseband precoding and analog RF precoding. The hybrid architecture can reduce the number of RF chains, while achieving a good trade-off between hardware complexity and performance. In addition, the passive reconfigurable intelligent surfaces (RIS) \cite{Huang2019} can be considered on the other end of the spectrum, where only the phase of the signals is controlled by the RIS elements with no active RF chains. The design of the beamforming/precoding for these architectures, however, requires knowledge of the channel state information (CSI), which is generally challenging to acquire due to the large number of antennas and the hardware constraints on the RF chains. This arises the need for efficient channel estimation methods to address the challenges of MIMO systems. In this paper, we focus on the channel estimation and hybrid precoding problem in MIMO systems.

    In the literature, several studies have proposed channel estimation and hybrid precoding methods for MIMO systems. Leveraging the sparsity of the channel, compressive sensing based channel estimation methods have been developed to reduce the overhead of channel estimation \cite{Alkhateeb2014,Schniter2014,Ding2018,Xie2020}. These methods quantize the channel with an over-complete dictionary and leverage random measurements to perform sparse recovery. Subsequently, hybrid precoding is designed based on the estimated channel. However, employing random measurements for channel estimation may not be efficient due to the unnecessary power distribution among all spatial directions. Recently, deep learning has demonstrated its capability in learning complex patterns from data and has been applied to jointly optimize the channel sensing vectors and design the hybrid precoding \cite{Li2019}. Compared to the traditional methods, deep learning based approaches can effectively learn the compressive sensing measurement vectors that capture promising spatial directions of the channel for a specific environment, resulting in less overhead in channel estimation. Data-driven approaches, however, require a large amount of data to achieve good performance. This leads to a high overhead in data collection, which is not practical in real-world systems.

    In this paper, we employ a site-specific digital twin \cite{Alkhateeb2023} to reduce the overhead of data collection in the deep learning based compressive sensing for hybrid precoding. Digital twin has demonstrated its capability in enhancing the performance of several tasks, such as beam prediction~\cite{Jiang2023}, channel compression and feedback~\cite{Jiang2024}, and localization~\cite{Morais2024}, by providing site-specific synthetic data. The contribution of this paper can be summarized as follows:
    \begin{itemize}
        \item We propose to leverage a site-specific digital twin to generate synthetic channel data, which shares the similar distribution as the real-world data. The synthetic data can be used to train the deep learning models for learning compressive sensing measurement vectors and predicting RF precoder/combiner in an end-to-end manner.
        \item We propose a model refinement approach to fine-tune the model pre-trained on the digital twin data with a small amount of real-world data. The performance can be further improved by mitigating the mismatch between the synthetic and real-world data distributions.
    \end{itemize}
    The evaluation results demonstrate the effectiveness of the proposed digital twin aided compressive sensing approach for hybrid precoding in reducing the overhead of data collection and achieving high performance on the real-world deployment.

\begin{figure*}
    \centering
    \includegraphics[width=0.9\textwidth]{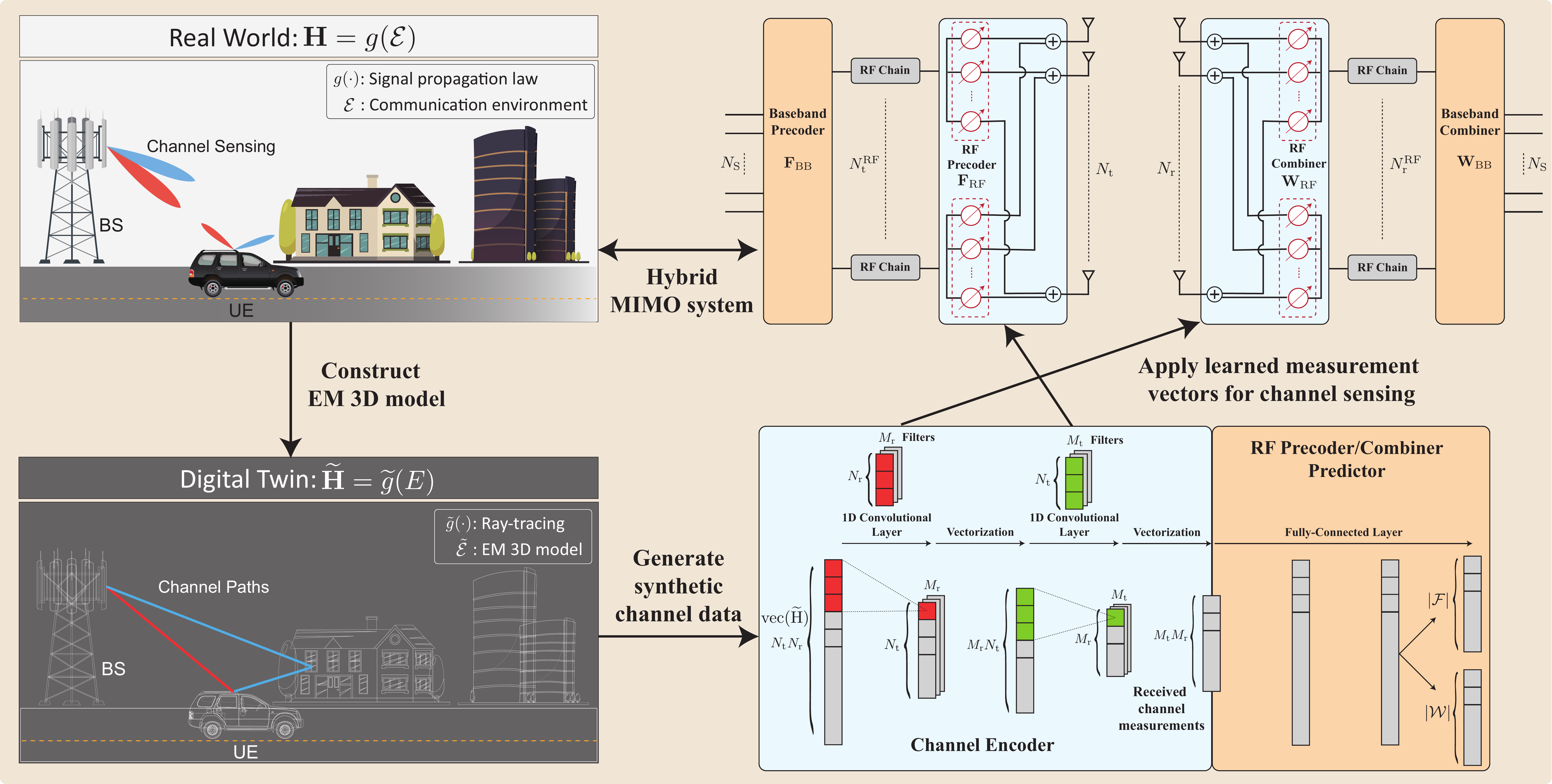}
    \caption{This diagram illustrates the adopted system model and the key idea of the proposed digital twin aided compressive sensing for hybrid precoding. The digital twin generates synthetic channel data, which shares a similar distribution as the real-world data. The channel encoder mimics the channel sensing process and learns the transmit and receive measurement vectors that capture the promising spatial directions of the channel. Based on the received channnel measurements, the RF precoder/combiner predictor predicts the codebook indices for the RF precoder/combiner. The RF precoder/combiner predictor, for instance, can be implemented in the baseband processor at the receiver.}
    \label{fig:sys_model}
\end{figure*}

\section{System Model}
    As shown in \figref{fig:sys_model}, we consider a massive MIMO system in which a transmitter, equipped with $N_{\rm{t}}$ antennas and $N_{\rm{t}}^{\rm{RF}}$ RF chains, communicates with a receiver employing $N_{\rm{r}}$ antennas and $N_{\rm{r}}^{\rm{RF}}$ RF chains. The transmitter precodes the transmit signal of $N_{\rm{S}}$ data streams using a hybrid precoder, where $N_{\rm{S}} \leq N_{\rm{t}}^{\rm{RF}} \leq N_{\rm{t}}$. The hybrid precoder comprises a baseband precoder $\bF_{\rm{BB}} \in \bbC^{N_{\rm{t}}^{\rm{RF}} \times N_{\rm{S}}}$ and an RF precoder $\bF_{\rm{RF}} \in \bbC^{N_{\rm{t}} \times N_{\rm{t}}^{\rm{RF}}}$. The receiver, then, combines the received signal using the RF combiner $\bW_{\rm{RF}} \in \bbC^{N_{\rm{r}} \times N_{\rm{r}}^{\rm{RF}}}$ and the baseband combiner $\bW_{\rm{BB}} \in \bbC^{N_{\rm{r}}^{\rm{RF}} \times N_{\rm{S}}}$. In this work, we assume the RF precoder and combiner are implemented using phase shifters, with their elements having a constant modulus, i.e., $\left|[\bF_{\rm{RF}}]_{m,n}\right|^2=N_{\rm{t}}^{-1}$ and $\left|[\bW_{\rm{RF}}]_{m,n}\right|^2=N_{\rm{r}}^{-1}$.

    \subsection{Signal Model}
        The transmitter precodes the transmit symbols using the hybrid precoder, and the discrete-time transmit signal $\bx \in \bbC^{N_{\rm{t}} \times 1}$ is given by
        \begin{align}
            \bx = \bF_{\rm{RF}} \bF_{\rm{BB}} \bs,
        \end{align}
        where $\bs \in \bbC^{N_{\rm{S}} \times 1}$ is the transmit symbol. The transmit symbol is supposed to have a power constraint, i.e., $\bbE[\bs \bs^H]=(P_{\rm{S}}/N_{\rm{S}})\bI_{N_{\rm{S}}}$, where $P_{\rm{S}}$ is the average total transmit power. To enforce the total transmit power constraint, the RF precoder and baseband precoder are designed such that $\|\bF_{\rm{RF}} \bF_{\rm{BB}}\|_F^2 = N_{\rm{S}}$. We adopt the narrowband block-fading channel model, where the received signal $\br \in \bbC^{N_{\rm{r}} \times 1}$ is given by
        \begin{align}
            \br = \bH \bF_{\rm{RF}} \bF_{\rm{BB}} \bs + \bn,
        \end{align}
        where $\bH \in \bbC^{N_{\rm{r}} \times N_{\rm{t}}}$ is the channel matrix, and $\bn \in \bbC^{N_{\rm{r}} \times 1}$ is the additive white Gaussian noise vector with $\bbE[\bn \bn^H]=\sigma^2\bI_{N_{\rm{r}}}$. The received signal is further processed by the RF and baseband combiners, which yields the processed received signal $\by \in \bbC^{N_{\rm{r}}^{\rm{RF}} \times 1}$ given by
        \begin{align}
            \by = \bW^H \bH \bF \bs + \bW^H \bn,
        \end{align}
        where $\bF = \bF_{\rm{RF}} \bF_{\rm{BB}}$ and $\bW = \bW_{\rm{RF}} \bW_{\rm{BB}}$. Then the spectral efficiency achieved by the system is given by
        \begin{align}
            R = \log_2 |\bI + \bQ^{-1} \bW^H \bH \bF \bF^H \bH^H \bW|,
        \end{align}
        where $\bQ = \frac{1}{\rm{SNR}} \bW^H \bW$ and $\rm{SNR} = \frac{P_{\rm{S}}}{N_{\rm{S}} \sigma^2}$.

    \subsection{Channel Model}
        We consider the geometric channel model, where the channel matrix $\bH$ is given by
        \begin{align}
            \bH = \sum_{l=1}^{L} \alpha_l \ba_{\rm{r}}(\theta_{l}) \ba_{\rm{t}}^H(\phi_{l}),
        \end{align}
        where $L$ is the number of paths, and $\alpha_l$ is the complex gain of the $l^{\rm{th}}$ path. $\phi_{l} \in [0, 2\pi]$ and $\theta_{l} \in [0, 2\pi]$ are the azimuth angles of departure or arrival (AoDs/AoAs) of the $l^{\rm{th}}$ path at the transmitter and receiver. $\ba_{\rm{r}}(.)$ and $\ba_{\rm{t}}(.)$ are the receive and transmit array response vectors, respectively. The channel matrix can be written in a more compact form as
        \begin{align} \label{eq:channel_approx}
            \bH = \bA_{\rm{r}} \textrm{diag}(\bm{\alpha}) \bA_{\rm{t}}^H,
        \end{align}
        where $\bm{\alpha}=\left[ \alpha_1, \ldots, \alpha_L \right]^T$, and the columns of $\bA_{\rm{r}}$ and $\bA_{\rm{t}}$ are the receive and transmit array response vectors considering the $L$ channel paths.

\section{Problem Formulation}
    In this work, we aim to learn the compressive sensing measurement vectors from the site-specific collected channel data, which can effectively capture the promising spatial directions of the channel with minimum redundancy. The measurement vectors are then used to sense the channel, and the hybrid precoder/combiner are designed based on the channel measurements. Let $\cD = \left\{ \bH_1, \ldots, \bH_{D} \right\}$ denote the collected channel data. To learn the measurement vectors, we define a learning function $f(\cdot)$, which takes the channel data as input and outputs the measurement vectors, given by
    \begin{align}
        \bP, \bQ = f(\cD),
    \end{align}
    where $\bP \in \bbC^{N_{\rm{t}} \times M_{\rm{t}}}$ and $\bQ \in \bbC^{N_{\rm{r}} \times M_{\rm{r}}}$ are the transmit and receive measurement matrices. During the training phase, if all pilot symbols are assumed to be the same~\cite{Alkhateeb2014}, the received measurement matrix $\bY \in \bbC^{M_{\rm{r}} \times M_{\rm{t}}}$ can be written as
    \begin{align} \label{eq:received_measurement}
        \bY = \sqrt{P} \bQ^H \bH \bP + \bQ^H \bN,
    \end{align}
    where $P$ is the average power used per transmission in the training phase. $\bN \in \bbC^{N_{\rm{r}} \times M_{\rm{t}}}$ is the noise matrix. To formulate the sparse recovery problem of the channel~\cite{Alkhateeb2014}, we first consider the vectorized received measurement matrix $\by = \rm{vec}(\bY)$, which can be written as
    \begin{align} \label{eq:vectorized_measurement}
        \by &= \sqrt{P} \rm{vec}( \bQ^H \bH \bP ) + \rm{vec}(\bQ^H \bN) \nonumber \\
        &= \sqrt{P} (\bP^T \otimes \bQ^H) \rm{vec}(\bH) + \bn_{\rm{Q}} \nonumber \\
        &= \sqrt{P} (\bP^T \otimes \bQ^H) \bA \bm{\alpha} + \bn_{\rm{Q}},
    \end{align}
    where $\bA \in \bbC^{N_{\rm{t}} N_{\rm{r}} \times L}$ is the Khatri-Rao product of $\bA_{\rm{r}}$ and $\bA_{\rm{t}}$. Then, the AoDs/AoAs of the channel are assumed be selected from a uniform grid of $N$ points, with $N \gg L$ and $\phi_l, \theta_l \in \left\{ 0, 2\pi/N, \ldots, 2\pi(N-1)/N \right\}, \forall l$. Thus, the vectorized received signal $\by$ can be approximated as
    \begin{align}
        \by = \sqrt{P} (\bP^T \otimes \bQ^H) \bA_{\rm{D}} \bz + \bn_{\rm{Q}},
    \end{align}
    where $\bA_{\rm{D}} \in \bbC^{N_{\rm{t}} N_{\rm{r}} \times N^2}$ is the dictionary matrix, and $\bz \in \bbC^{N^2 \times 1}$ contains the complex gains corresponding to the quantized directions. Since $\bz$ has only $L$ non-zero elements and $L \ll N^2$, the formulation of the received signal $\by$ can be seen as a sparse formulation of the channel estimation problem. This further implies that the number of measurements $M_{\rm{t}}$ and $M_{\rm{r}}$ required to estimate $\bz$ is much smaller than $N^2$. With the estimated $\widehat{\bz}$, the vectorized channel can be recovered as $\widehat{\bh} = \bA_{\rm{D}} \widehat{\bz}$, and the channel matrix $\widehat{\bH}$ can be obtained by reshaping $\widehat{\bh}$ to the matrix form.
    
    To reduce the design complexity, we assume the RF precoder/combiner are selected from pre-defined codebooks, i.e., $\bF_{\rm{RF}} \in \bm{\mathcal{F}}$ and $\bW_{\rm{RF}} \in \bm{\mathcal{W}}$. The hybrid precoder/combiner design problem can be formulated as
    \begin{equation} 
        \begin{aligned}
            \max_{\{\bF_{\rm{BB}}, \bF_{\rm{RF}}, \bW_{\rm{BB}}, \bW_{\rm{RF}}\}}  & \log_2 |\bI + \bQ^{-1} \bW^H \widehat{\bH} \bF \bF^H \widehat{\bH}^H \bW|,  \\
            \textrm{s.t.} \quad & \bF = \bF_{\rm{RF}} \bF_{\rm{BB}}, \\
            \quad & \bW = \bW_{\rm{RF}} \bW_{\rm{BB}}, \\
            \quad & \bF_{\rm{RF}} \in \bm{\mathcal{F}}, \ \ \forall n_{\rm{t}}, \\
            \quad & \bW_{\rm{RF}} \in \bm{\mathcal{W}}, \ \ \forall n_{\rm{r}}, \\
            \quad & \|\bF_{\rm{RF}} \bF_{\rm{BB}}\|_F^2 = N_{\rm{S}}. \\
        \end{aligned}
    \end{equation}
    If the RF precoder/combiner contain orthogonal vectors, then the optimal baseband precoder/combiner for the selected RF precoder/combiner can be obtained as \cite{Alkhateeb2016}
    \begin{equation}
        \bF_{\rm{BB}}^\star = \left( \bF_{\rm{RF}}^H \bF_{\rm{RF}} \right)^{-\frac{1}{2}} \left[ \, \overline{\bV} \,\right]_{:, 1:N_{\rm{S}}}, \\
    \end{equation}
    \begin{equation}
        \bW_{\rm{BB}}^\star = \left[ \, \overline{\bU} \, \right]_{:, 1:N_{\rm{S}}},
    \end{equation}
    where $\overline{\bV}$ and $\overline{\bU}$ are the left singular matrices of the effective channel matrix $\overline{\bH} = \bW_{\rm{RF}}^H \widehat{\bH} \bF_{\rm{RF}}$. Thus, the hybrid precoder/combiner design problem can be simplified to the search problem of the optimal RF precoder/combiner, which can be formulated as 
    \begin{equation} \label{eq:hybrid_precoding}
        \begin{aligned}
            \max_{\{\bF_{\rm{RF}}, \bW_{\rm{RF}}\}}  & \log_2 |\bI + \rm{SNR} \, \bW_{\rm{RF}}^H \widehat{\bH} \bF_{\rm{RF}} \\
            & \quad \times \left( \bF_{\rm{RF}}^H \bF_{\rm{RF}} \right)^{-\frac{1}{2}} \bF_{\rm{RF}}^H \widehat{\bH}^H \bW_{\rm{RF}}|,  \\
            \textrm{s.t.} \quad & \bF_{\rm{RF}} \in \bm{\mathcal{F}}, \ \ \forall n_{\rm{t}}, \\
            \quad & \bW_{\rm{RF}} \in \bm{\mathcal{W}}, \ \ \forall n_{\rm{r}}, \\
        \end{aligned}
    \end{equation}
    where the optimal solution can be obtained by exhaustive search over the RF precoding/combining codebooks.

    \textbf{Main challenges:} The main challenges in the proposed problem formulation are as follows: (i) The objective of learning the compressive sensing measurement vectors is to leverage the minimum number of measurements to capture the promising spatial directions of the channel. Defining this objective, however, is non-trivial. (ii) The performance of the problem in \eqref{eq:hybrid_precoding} is highly dependent on the quality of the estimated channel, which is affected by the quality of the learned measurement vectors. To address these challenges, we adopt a deep learning based compressive sensing approach~\cite{Li2019}, where the measurement vectors are learned in an unsupervised manner, and the end objective is to optimize the prediction of RF precoder/combiner. In particular, the number of measurements is a hyperparameter of the deep learning model, which can be optimized based on the performance of the RF precoder/combiner prediction. Moreover, to reduce the overhead of data collection in the real-world systems, we propose to leverage a site-specific digital twin to generate the training data for the deep learning model, which will be discussed in the next section.

\section{Proposed Solutions}
    In this section, we first introduce the main concept of using the digital twin. Then, we present the deep learning based compressive sensing for hybrid precoding. Finally, we propose the digital twin aided compressive sensing approach.

    \subsection{Key Idea}
        The objective of learning compressive sensing measurement vectors is to identify the promising spatial directions of the channel $\bH$, which can be determined by the communication environment $\cE$ and the wireless propagation law $g(\cdot)$, i.e., $\bH = g(\cE)$. Thus, learning measurement vectors is highly dependent on the environment in which the system operates. However, collecting real-world channel data is challenging in general. To reduce the data collection overhead, we propose to leverage a site-specific digital twin to generate the synthetic data for learning compressive sensing measurement vectors. The digital twin~\cite{Jiang2023} can be used to approximate the communication environment $\cE$ and the wireless propagation law $g(\cdot)$ by using the EM 3D model $E$ and the ray tracing model $\widetilde{g}(\cdot)$. Accordingly, the synthetic channel data can be generated as $\widetilde{\bH}=\widetilde{g}(E)$. Then, site-specific synthetic data, which has the similar distribution as the real-world data, can be employed in the training process and achieve high performance on the real-world data.

    \subsection{Deep Learning Based Compressive Sensing for Hybrid Precoding}
        We adopt a deep learning based approach~\cite{Li2019} to learn the compressive sensing measurement vectors for the hybrid precoding in an end-to-end manner. The deep learning model contains two parts: Channel encoder and RF precoder/combiner predictor.
        
        \textbf{Channel Encoder:}
        The channel encoder aims to mimic the channel estimation process by learning the transmit and receive measurements, i.e., $\bP$ and $\bQ$. Specifically, the Kronecker product in \eqref{eq:vectorized_measurement} can be emulated by passing the channel vector $\bh$ through two consecutive 1D convolutional layers. The first convolutional layer contains $M_{\rm{r}}$ filters, where each filter is of size $N_{\rm{r}}$ and the stride is set to $N_{\rm{r}}$. The second convolutional layer contains $M_{\rm{t}}$ filters, where each filter is of size $N_{\rm{t}}$ and the stride is set to $N_{\rm{t}}$. It is worth noting that, after the training, the weights of the convolutional layers can be adopted by the transmitter and receiver to measure the unknown channel.
        
        \textbf{RF Precoder/Combiner Predictor:}
        The output of the channel encoder, i.e., received channel measurements, is then fed into two sets of fully connected layers, which predict the beam indices for the RF precoder and combiner, respectively. Let $f_{\rm{enc}}(\cdot)$, $f_{\rm{pred},\rm{t}}(\cdot)$, and $f_{\rm{pred},\rm{r}}(\cdot)$ denote the channel encoder, RF precoder predictor, and RF combiner predictor, respectively. The whole process can be formulated as
        \begin{align}
            \widehat{\bp} &= f_{\rm{pred},\rm{t}}(f_{\rm{enc}}(\bh)), \\
            \widehat{\bq} &= f_{\rm{pred},\rm{r}}(f_{\rm{enc}}(\bh)),
        \end{align}
        where $\widehat{\bp} \in \bbR^{|\cF| \times 1}$ and $\widehat{\bq} \in \bbR^{|\cW| \times 1}$ are the probability distributions of the indices for the RF precoder and combiner, respectively.
        
        \textbf{Objective Function:}
        The loss function is defined as the cross entropy between the predicted and optimal indices, given by
        \begin{equation}
            {\rm{CE}} (\bp, \bq, \widehat{\bp}, \widehat{\bq}) = - \left(\sum_{i=1}^{|\cF|} p_i^\star \log \widehat{p}_i + \sum_{j=1}^{|\cW|} q_j^\star \log \widehat{q}_j\right),
        \end{equation}
        where $p_i^\star$ and $q_j^\star$ are the one-hot vectors of the optimal RF precoder and combiner. Then, the optimal model parameters can be obtained by minimizing the expected cross entropy loss, given by
        \begin{align}
            &f_{\rm{enc}}^\star(;\Theta_{\rm{enc}}^\star), f_{\rm{pred},\rm{t}}^\star(;\Theta_{\rm{pred},\rm{t}}^\star), f_{\rm{pred},\rm{r}}^\star(;\Theta_{\rm{pred},\rm{r}}^\star) \nonumber \\
            = &\argmin_{\substack{f_{\rm{enc}}(;\Theta_{\rm{enc}}) \\ f_{\rm{pred},\rm{t}}(;\Theta_{\rm{pred},\rm{t}}) \\ f_{\rm{pred},\rm{r}}(;\Theta_{\rm{pred},\rm{r}})}} \bbE_{\bh \sim \cH} \left\{ {\rm{CE}}(\bp, \bq, \widehat{\bp}, \widehat{\bq}) \right\} ,
        \end{align}
        where $\Theta_{\rm{enc}} = \left\{ \bP, \bQ \right\}$, $\Theta_{\rm{pred},\rm{t}}$, and $\Theta_{\rm{pred},\rm{r}}$ are the model parameters of the channel encoder, RF precoder predictor, and RF combiner predictor, respectively. Typically, a dataset of the channel data is used to approximate the exact channel distribution, and the model parameters are trained by minimizing the following loss function
        \begin{align} \label{eq:deep_learning_loss}
            &\cL(\Theta_{\rm{enc}}, \Theta_{\rm{pred},\rm{t}}, \Theta_{\rm{pred},\rm{r}}, \cD) \nonumber \\
            = &\frac{1}{D} \sum_{d=1}^{D} {\rm{CE}} (\bp, \bq, f_{\rm{enc}}(f_{\rm{pred},\rm{t}}(\bh_d)), f_{\rm{enc}}(f_{\rm{pred},\rm{r}}(\bh_d))),
        \end{align}
        where $\cD = \left\{ \bh_1, \ldots, \bh_D \right\}$ is the dataset of the channel data.

    \subsection{Digital Twin Aided Compressive Sensing for Hybrid Precoding}
        In this work, we propose to leverage the site-specific digital twin to generate the synthetic channel data for the deep learning model. By training the model on synthetic data, our objective is to achieve similar performance as the model trained on real-world data. Let $\widetilde{f}_{\rm{enc}}(;\widetilde{\Theta}_{\rm{enc}})$, $\widetilde{f}_{\rm{pred},\rm{t}}(;\widetilde{\Theta}_{\rm{pred},\rm{t}})$, and $\widetilde{f}_{\rm{pred},\rm{r}}(;\widetilde{\Theta}_{\rm{pred},\rm{r}})$ denote the channel encoder, RF precoder/combiner predictors trained on the distribution in the digital twin $\widetilde{\cH}$. The objective can be formulated as
        \begin{align}
            \min_{\substack{\widetilde{f}_{\rm{enc}}(;\widetilde{\Theta}_{\rm{enc}}) \\ \widetilde{f}_{\rm{pred},\rm{t}}(;\widetilde{\Theta}_{\rm{pred},\rm{t}}) \\ \widetilde{f}_{\rm{pred},\rm{r}}(;\widetilde{\Theta}_{\rm{pred},\rm{t}})}} &\Big| \cL \left( \widetilde{\Theta}_{\rm{enc}}, \widetilde{\Theta}_{\rm{pred},\rm{t}}, \widetilde{\Theta}_{\rm{pred},\rm{r}}, \cH \right) \nonumber \\ 
            &- \cL \left( \Theta_{\rm{enc}}^\star, \Theta_{\rm{pred},\rm{t}}^\star, \Theta_{\rm{pred},\rm{r}}^\star, \cH \right) \Big|,
        \end{align}
        where $\cH$ is the real-world channel distribution. To achieve this objective, we propose two approaches, i.e., direct generalization and model refinement.

        \textbf{Direct Generalization:}
        Digital twin aided deep learning can be seen as a domain adaptation problem~\cite{Jiang2024}, where the source domain is the synthetic data distribution and the target domain is the real-world data distribution. By directly applying the model trained on the synthetic data to the real-world data, the domain adaptation bound~\cite{Mansour2009} can be obtained as
        \begin{align} \label{eq:domain_adaptation}
            &\Big| \cL \! \left( \widetilde{\Theta}_{\rm{enc}}, \widetilde{\Theta}_{\rm{pred},\rm{t}}, \widetilde{\Theta}_{\rm{pred},\rm{r}}, \cH \right) \! -  \! \cL \! \left( \Theta_{\rm{enc}}^\star, \Theta_{\rm{pred},\rm{t}}^\star, \Theta_{\rm{pred},\rm{r}}^\star, \cH \right) \! \Big| \nonumber \\ 
            &\leq  \cL\left( \widetilde{\Theta}_{\rm{enc}}, \widetilde{\Theta}_{\rm{pred},\rm{t}}, \widetilde{\Theta}_{\rm{pred},\rm{r}}, \widetilde{\cH} \right) + {\rm{disc}} \left( \cH, \widetilde{\cH} \right) + \epsilon,
        \end{align}
        where $\rm{disc}(\cdot)$ denotes the discrepancy between two distributions, and $\epsilon$ is a constant determined by the data distributions and the function class of the model. From \eqref{eq:domain_adaptation}, we can observe that both $\rm{disc}(\cdot)$ and $\epsilon$ are constants given the distributions $\cH$ and $\widetilde{\cH}$, and the upper bound of the objective can be minimized by reducing $\cL\left( \widetilde{\Theta}_{\rm{enc}}, \widetilde{\Theta}_{\rm{pred},\rm{t}}, \widetilde{\Theta}_{\rm{pred},\rm{r}}, \widetilde{\cH} \right)$, i.e., the loss on the synthetic data distribution. Thus, we can improve the performance of the model on the real-world data by minimizing the following loss function
        \begin{align}
            &\cL\left( \widetilde{\Theta}_{\rm{enc}}, \widetilde{\Theta}_{\rm{pred},\rm{t}}, \widetilde{\Theta}_{\rm{pred},\rm{r}}, \widetilde{\cD} \right) \nonumber \\
            = &\frac{1}{\widetilde{D}} \sum_{d=1}^{\widetilde{D}} {\rm{CE}} (\bp, \bq, f_{\rm{enc}}(f_{\rm{pred},\rm{t}}(\bh_d)), f_{\rm{enc}}(f_{\rm{pred},\rm{r}}(\bh_d))),
        \end{align}
        where $\widetilde{\cD} = \left\{ \widetilde{\bh}_1, \ldots, \widetilde{\bh}_{\widetilde{D}} \right\}$ is the dataset of the synthetic channel data.

        \textbf{Model Refinement:}
        In practice, constructing a digital twin that perfectly mimics the real-world environment is challenging, which may result in a degraded performance due to a large $\rm{disc} (\cH, \widetilde{\cH})$. To mitigate this issue, we propose to refine the model trained on the synthetic data by fine-tuning it on a small amount of real-world data $\cD_{\rm{r}}$. The real-world data, for instance, can be collected from the feedback provided the UEs during the online operations~\cite{Jiang2024}. For the fine-tuning strategy, we adopt the rehearsal approach~\cite{Robins1995}, where the model is trained on both previously learned data samples and new data samples. Rehearsal helps prevent the model from forgetting the previously learned knowledge from the digital twin synthetic data. Consequently, the model is refined by minimizing the loss function described in \eqref{eq:deep_learning_loss} on the combined dataset $\cD = \cD_{\rm{r}} \cup \widetilde{\cD}$.

\begin{figure}
    \centering
    \includegraphics[width=0.48\textwidth]{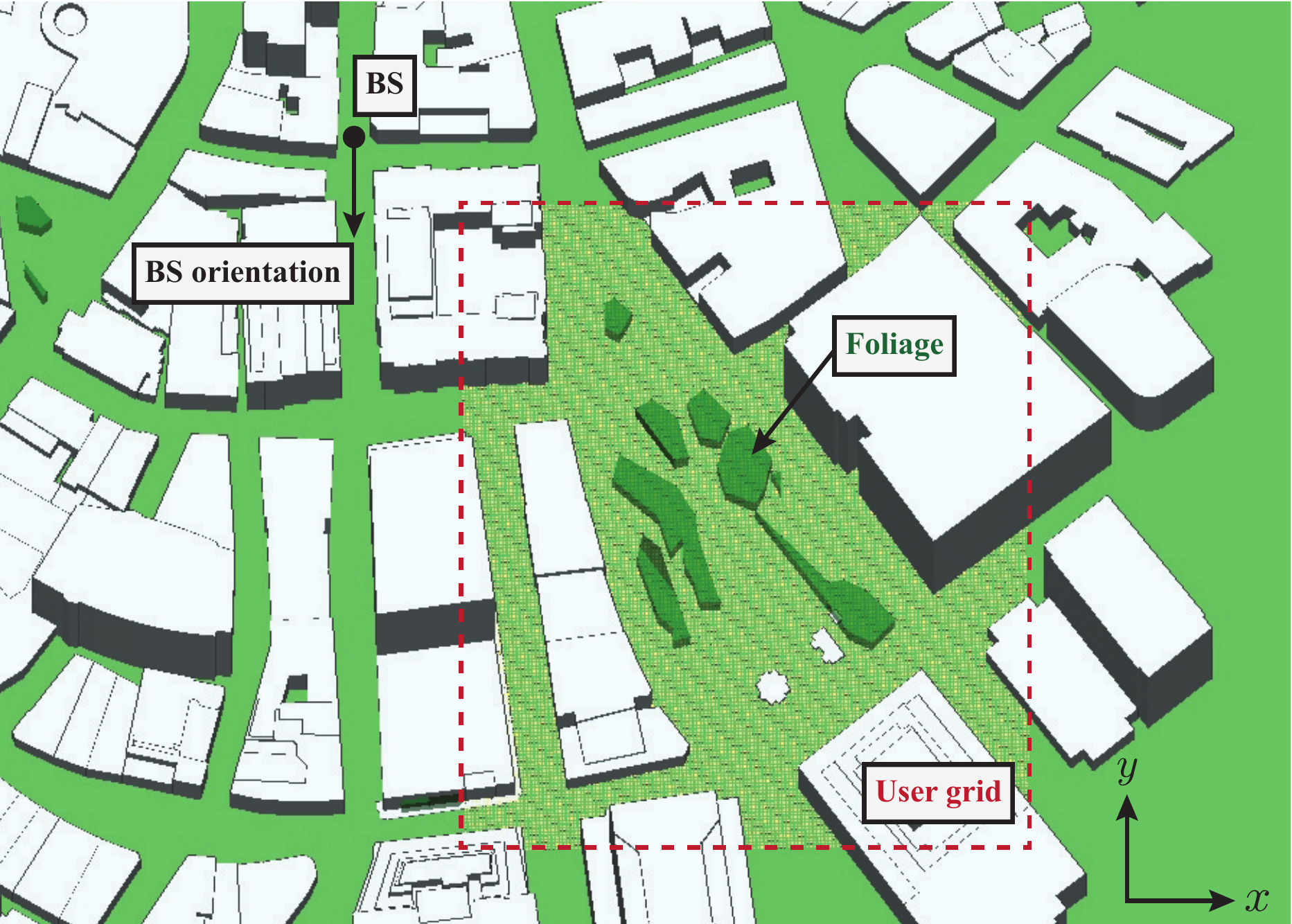}
    \caption{The figure shows the adopted target scenario, which is built based on a section of downtown Boston. The base station is placed along a vertical street, while the foliage is represented by the dark green objects in the layout. The user grid is highlighted by the red box.}
    \label{fig:target_scenario}
\end{figure}

\section{Simulation Setup}
    In this section, we present the simulation setup, data generation, and deep learning model architecture.
    \subsection{Scenario Setup}
        \textbf{Target Scenario:}
        The target scenario represents real-world deployment in the simulation. As shown in Fig.~\ref{fig:target_scenario}, we adopt an urban scenario, which is constructed based on a section of downtown Boston. This scenario consists of a BS, a service area, buildings, and foliage objects. The BS is assumed to employ a uniform linear array (ULA) with $N_{\rm{t}}=32$ antennas and $N_{\rm{t}}^{\rm{RF}}=1$ RF chains. It is placed at a height of $15$ meters and oriented towards the negative $y$-axis. The BS communicates with a single-antenna UE, i.e., $N_{\rm{r}}=N_{\rm{r}}^{\rm{RF}}=1$, which is located in the service area at a height of $2$ meters. The scenario of the multiple-antenna UE is left for future work. The service area measures $200$ meters by $230$ meters and is discretized into a user grid with a spacing of $0.37$ meters.

        \textbf{Digital Twin Scenario:}
        In practice, building a digital twin that perfectly represents the real-world environment is challenging, and there may be some modeling errors. For instance, accurately modeling foliage is difficult due to seasonal changes and the randomness of its growth patterns. Additionally, the precise positions of buildings may not be known. Therefore, the adopted digital twin scenario shares the same layout as the target scenario but with some modifications. Foliage objects are neglected entirely, and the buildings are randomly moved from their real-world positions on the X-Y plane. In \sref{sec:evaluation}, we will study the impact of these modeling errors on the performance of the hybrid precoding design.

    \subsection{Dataset Generation} 
        We utilize Wireless Insite~\cite{Remcom} to conduct ray tracing simulations in both the target and digital twin scenarios. The ray tracing simulations operates at a carrier frequency of $3.5$ GHz, and the propagation paths between the BS and UE are searched up to the $4^{\rm{th}}$ order reflection. For each path, the complex gain $\alpha_l$ and the azimuth AoDs/AoAs $\phi_{l}$, $\theta_{l}$ are obtained. Then, we use DeepMIMO generator~\cite{Alkhateeb2019} to construct the channel between the BS and UE. Optimal RF precoder for each channel is obtained by exhaustive search over a 32-element DFT codebook. The pair of channel and the codebook index of the RF precoder is then considered as one data point. The resulting data points are then partitioned into training and testing sets with an $80\%$ to $20\%$ ratio, respectively.

    \subsection{Deep Learning Model Architecture} 
        Since we consider a single-antenna UE, the deep learning model is designed to predict the RF precoder only. The channel encoder is implemented by a 1D complex-valued convolutional layer for the transmitter measurement vectors. The RF precoder predictor is implemented by two fully connected layers and an output layer, and each fully connected layer is followed by a sigmoid activation function. After each weight update, the convolutional filters are normalized to satisfy the constraints of RF precoder.

\section{Evaluation Results} \label{sec:evaluation}
    In this section, we evaluate the performance of the proposed digital twin aided compressive sensing for hybrid precoding.

    \subsection{Does the model trained on the digital twin data work in the real-world deployment?}
\begin{figure}
            \centering
            \includegraphics[width=0.49\textwidth]{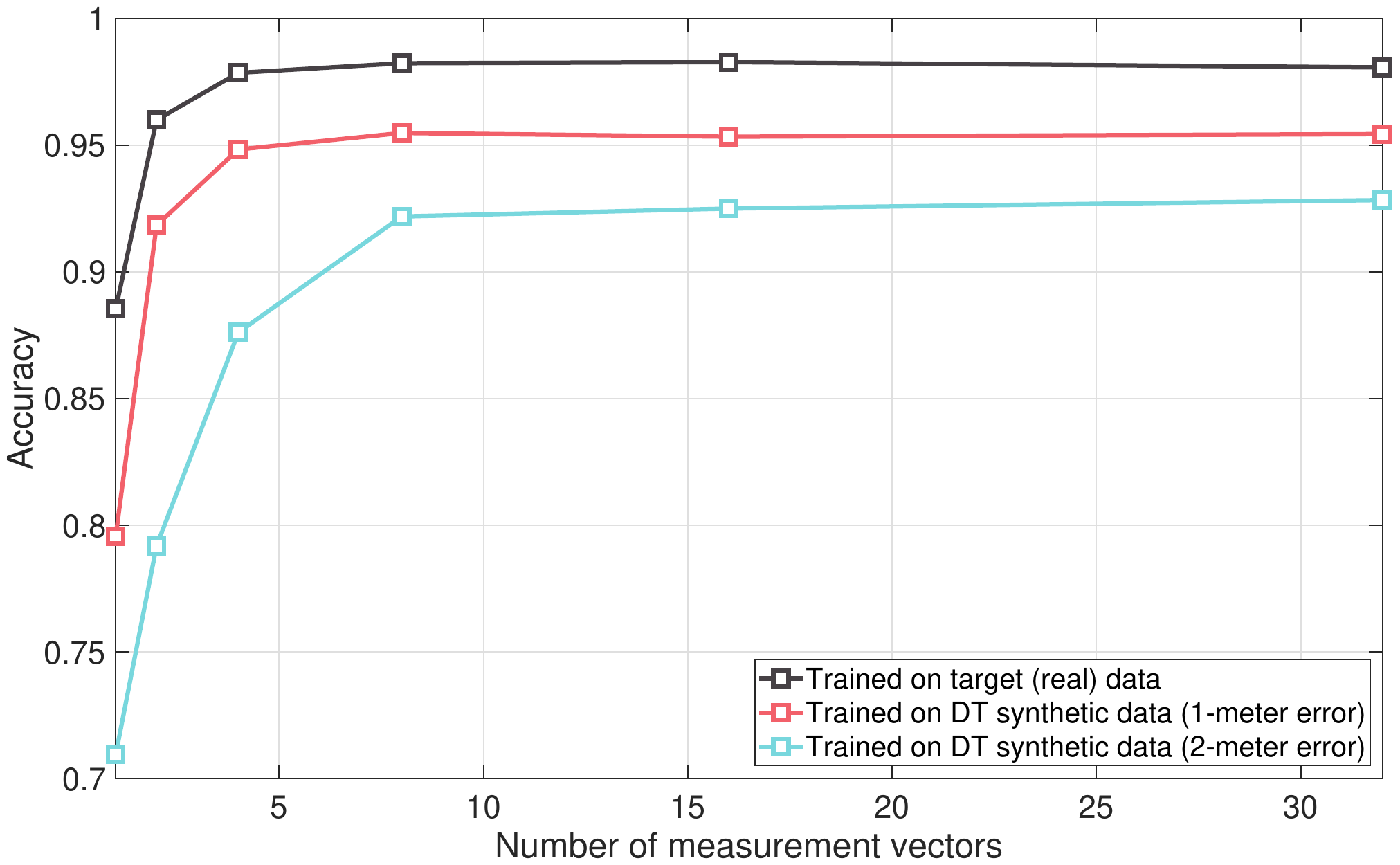}
            \caption{This figure presents the prediction accuracy of the RF precoder with different numbers of measurement vectors. We consider the number of measurement vectors to be set to $\{1,2,4,8,16,32\}$.}
            \label{fig:acc_vs_dict_size}
        \end{figure}

        We first evaluate the performance of the digital twin aided compressive sensing in real-world deployment. In Fig.~\ref{fig:acc_vs_dict_size}, we present the prediction accuracy of the RF precoder with different number of measurement vectors. We compare the models trained on digital twin data and target data, using $10240$ data points for training. All models are then tested on the unseen target data. As the number of measurement vectors increases, the prediction accuracy improves since the measurement vectors can capture more spatial directions for channel sensing. Also, we observe that, despite modeling errors in the digital twin, the model trained on synthetic data can achieve high performance on target data. For instance, with a modeling error of $1$ meter in building positions, the model can achieve an accuracy of $95\%$. It is worth noting that the model has only seen the synthetic data during the training phase, without any real-world data. The performance degrades as the modeling errors increase due to the mismatch between the synthetic and target data distributions. In addition, we observe $8$ measurement vectors is sufficient to capture the spatial directions of the channel in the considered scenario. This reveals that only $8$ channel measurements are needed, which is much less than $32$ pilots required in the exhaustive search method.

        \begin{figure*}
            \centering
            
            \subfigure[Trained on target (real) data]{\label{fig:beam_real}\includegraphics[width=0.49\textwidth]{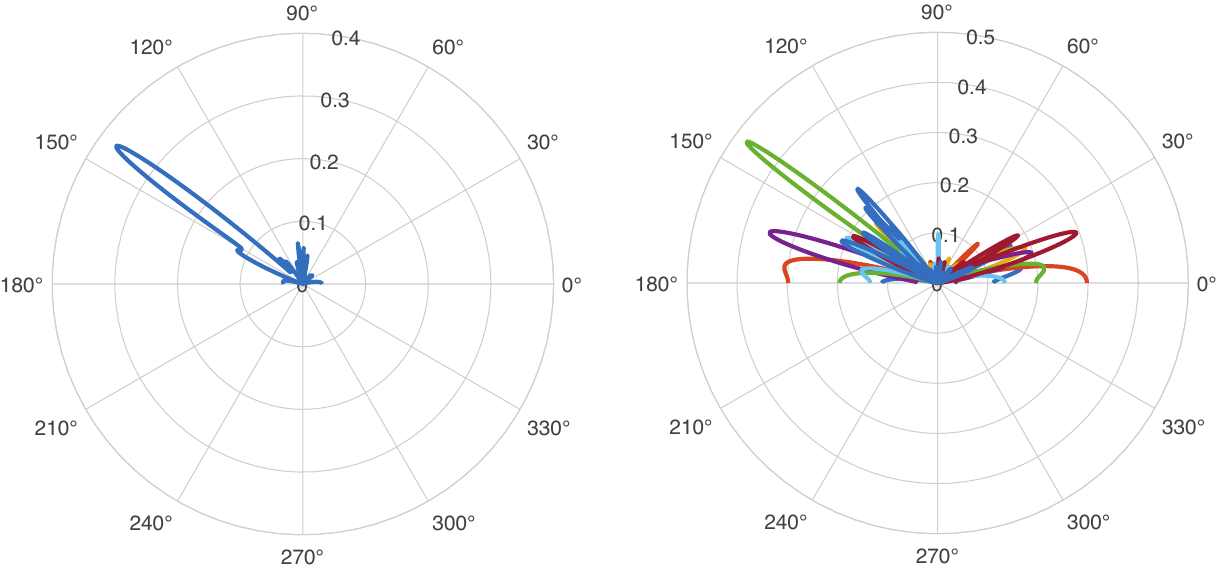}}
            \subfigure[Trained on digital twin synthetic data (1-meter error)]{\label{fig:beam_DT}\includegraphics[width=0.49\textwidth]{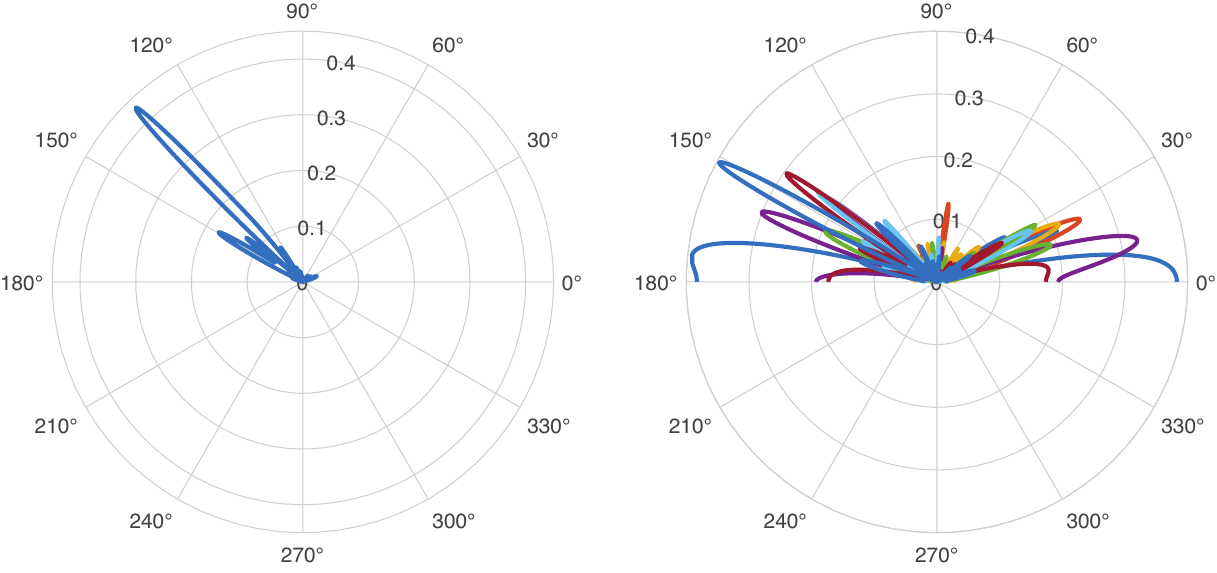}}
    
            \caption{This figure shows the beam patterns of the measurement vectors learned by the channel encoder. The number of measurement vectors is set to $\{1,8\}$. Fig.~\ref{fig:beam_real} shows the learned beam patterns by the model trained on the target data, while Fig.~\ref{fig:beam_DT} presents the learned beam patterns by the model trained on the digital twin data.}
            \label{fig:beam_pattern}
        \end{figure*}
        
    \subsection{Can the measurement vectors be learned from the digital twin?}
        Next, we investigate the beam patterns of the measurement vectors by the channel encoder. In particular, we are interested in whether the learned measurement vectors can be adapted to the environment. In Fig.~\ref{fig:beam_real}, we present the beam patterns learned by the model trained on the target data. We can observe that the learned measurement vectors focus the sensing power on the directions ranging from $120^\circ$ to $180^\circ$. This is consistent with the fact that the channel paths between the BS and UE mainly come from the left side of the BS. In Fig.~\ref{fig:beam_DT}, we present the beam patterns learned by the model trained on the digital twin data. The learned measurement vectors are similar to the target data, indicating that the model can capture the promising spatial directions of the channel from the synthetic data. This result demonstrates that the digital twin can be used to generate the synthetic data for the deep learning model, which can achieve high performance on the real-world data.     

    \subsection{Does the model refinement improve the performance?}
\begin{figure}
            \centering
            \includegraphics[width=0.49\textwidth]{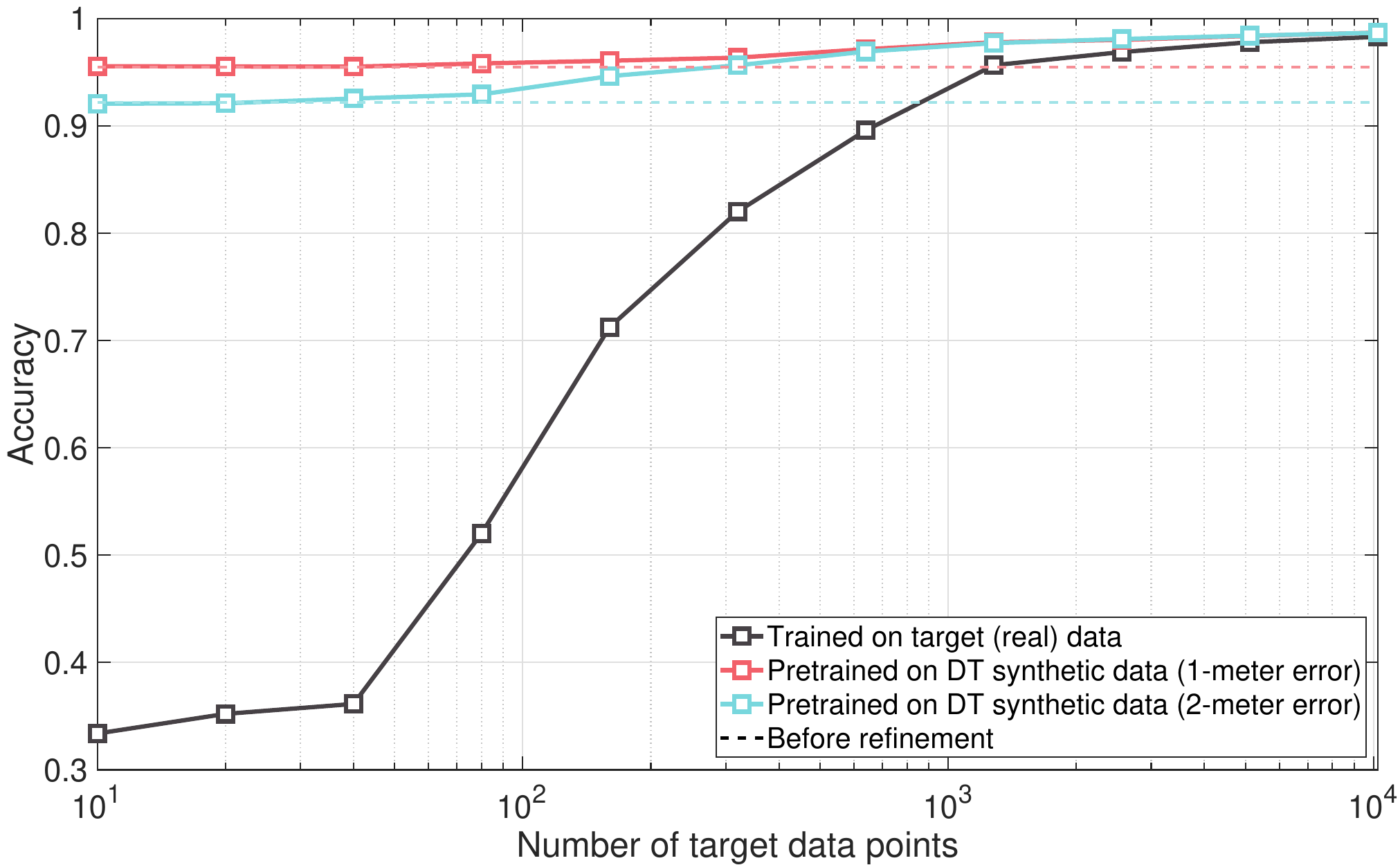}
            \caption{This figure presents the prediction accuracy of the RF precoder with different numbers of real-world data points. The model is pretrained on $10240$ synthetic data points and then fine-tuned on real-world data.}
            \label{fig:acc_vs_num_target_data}
        \end{figure}

        Finally, we evaluate the performance of the model refinement approach, where the number of measurement vectors is set to $8$. The model is first pre-trained on $10240$ synthetic data points and then fine-tuned on real-world data. In Fig.~\ref{fig:acc_vs_num_target_data}, we present the prediction accuracy of the RF precoding vector with different numbers of real-world data points. With the refining on real-world data, the model pre-trained on digital twin synthetic data can be further improved. Specifically, to achieve the same performance as the model trained on target data, less number of real-world data points are needed. For instance, the model pre-trained in digital twin with $1$ meter modeling error can achieve an accuracy of $98\%$ with only $2560$ real-world data points, which is $4$ times less than the model trained on target data. This result shows that, with the aid of the digital twin, the overhead of data collection in the real-world systems can be significantly reduced.

\section{Conclusion}
    In this paper, we explore the use of a digital twin in the deep learning based compressive sensing for hybrid precoding. Specifically, to reduce the overhead of data collection in real-world systems, we propose to leverage a site-specific digital twin to generate the synthetic data for the training process. The site-specific digital twin is constructed based on the EM 3D model and the ray tracing model, which can approximate the real-world communication environments. The generated synthetic data has the similar distribution as the real-world data, and the deep learning model is trained using this synthetic dataset. Further, we propose a model refinement approach, where the model is fine-tuned on a small amount of real-world data to mitigate the mismatch between the synthetic and target data distributions. The evaluation results show that the model trained on the digital twin data can achieve high performance on the real-world data, in terms of the prediction accuracy of the RF precoder. Moreover, the model refinement approach can further improve the performance of the model pre-trained on the synthetic data with fewer real-world data points compared to the model trained solely on target data.

\end{document}